\documentclass[
floats,eqsecnum,aps,epsf,graphicx,twocolumn ]{revtex4}

\usepackage{epsfig}

\catcode`\@=11

\begin{document}
\title{
What is the origin of chirality in the cholesteric phase of virus
suspensions ?}
\author{Eric Grelet}
\altaffiliation[Permanent address: ]{
 Centre de Recherche Paul Pascal, UPR 8641,
Avenue Albert Schweitzer, F-33600 Pessac, France.}
\author{Seth Fraden}
\affiliation{Department of Physics, MS-057, Brandeis University,
Waltham, MA 02454, U.S.A.}

\date{\today}

\begin{abstract}

We report a study of the cholesteric phase in monodisperse
suspensions of the rod-like virus {\it fd} sterically stabilized
with the polymer polyethylene glycol (PEG). After coating the
virus with neutral polymers, the phase diagram and nematic order
parameter of the {\it fd}-PEG system then become independent of
ionic strength. Surprisingly, the {\it fd}-PEG suspensions not
only continue to exhibit a cholesteric phase, which means that the
grafted polymer does not screen all chiral interactions between
rods, but paradoxically the cholesteric pitch of this sterically
stabilized {\it fd}-PEG system varies with ionic strength.
Furthermore, we observe that the cholesteric pitch decreases with
increasing viral contour length, in contrast to theories which
predict the opposite trend. Different models of the origin of
chirality in colloidal liquid crystals are discussed.

\end{abstract}


\maketitle \preprint{HEP/123-qed}

Chirality leads to a wonderful variety of liquid crystalline
phases \cite{deGennes}, from the cholesteric state to very complex
structures such as smectic blue phases \cite{BPSm}. In spite of
its importance, the connection between chirality at the molecular
scale and the macroscopic chiral structure of liquid crystalline
phases is not yet understood, and quantitative prediction of the
twist periodicity, also called cholesteric pitch, based on
molecular features
remains a 
unsolved problem.  
The elaboration of a rigorous statistical theory for the chiral
interaction is one of the greatest challenges in the physics of
liquid crystals. 

We report and discuss experimental studies of the cholesteric
phase of colloidal suspensions of filamentous bacteriophages,
denoted {\it{fd}}. The rod-like virus {\it{fd}} is a micron-length
semi-flexible polyelectrolyte formed by a single stranded DNA
around which coat proteins are helicoidally wrapped. 
Therefore {\it{fd}} is a chiral
object exhibiting a helical charge distribution. Suspensions of
{\it{fd}} rods in aqueous solution form several liquid crystalline
phases with increasing concentration, and especially a cholesteric
or chiral nematic phase which lies between the isotropic and
smectic phases \cite{PRLDogic}. However, molecular chirality does
not guaranty the existence of macroscopic chiral structure such as
a cholesteric phase. For instance other chiral viruses such as
Tobacco Mosaic Virus (TMV) \cite{TMV}, or Pf1 \cite{langmuir} with
a helical structure extremely similar to {\it fd}, exhibit only a
nematic phase. 
Up to now, theory has failed to answer even the simple qualitative
question:
When does macroscopic chirality, such as twist, occur in liquid
crystals formed by
helical molecules? 
In this Letter, by studying a well characterized 
system of rod-like virus stabilized by grafted neutral polymer, we
highlight that counterions play a crucial role. More specifically
we show that ionic strength is one of the major physical
parameters controlling the chirality of colloidal liquid
crystalline suspensions.


The bacteriophage {\it fd} is a molecule with a contour length of
L=880~nm, a diameter of $D_{\mbox{\scriptsize bare}}$=6.6~nm, a
persistence length of $\xi$=2200 nm, a molecular weight of
M$_W$=1.64$\times$10$^7$ g/Mol, and a charge density of 10 e/nm in
water at pH=8.1 \cite{langmuir}. About 2700 copies of a single
protein form the coat helicoidally wrapped following a 5-fold
rotation axis combined with an 
2-fold screw axis around the inner DNA \cite{makowski}.
The {\it fd} virus was grown and purified following standard
biological protocols \cite{maniatis}. To vary ionic strength (I),
viruses were extensively dialyzed against a 20 mM TRIS-HCl buffer
at pH=8.1 with an adjusted amount of NaCl. All measurements were
done at room temperature and the concentrations were determined
using spectrophotometry with absorption coefficient of 3.84
cm$^2$/mg at 269 nm.

\begin{figure}[tbh]
\centerline{\epsfig{file=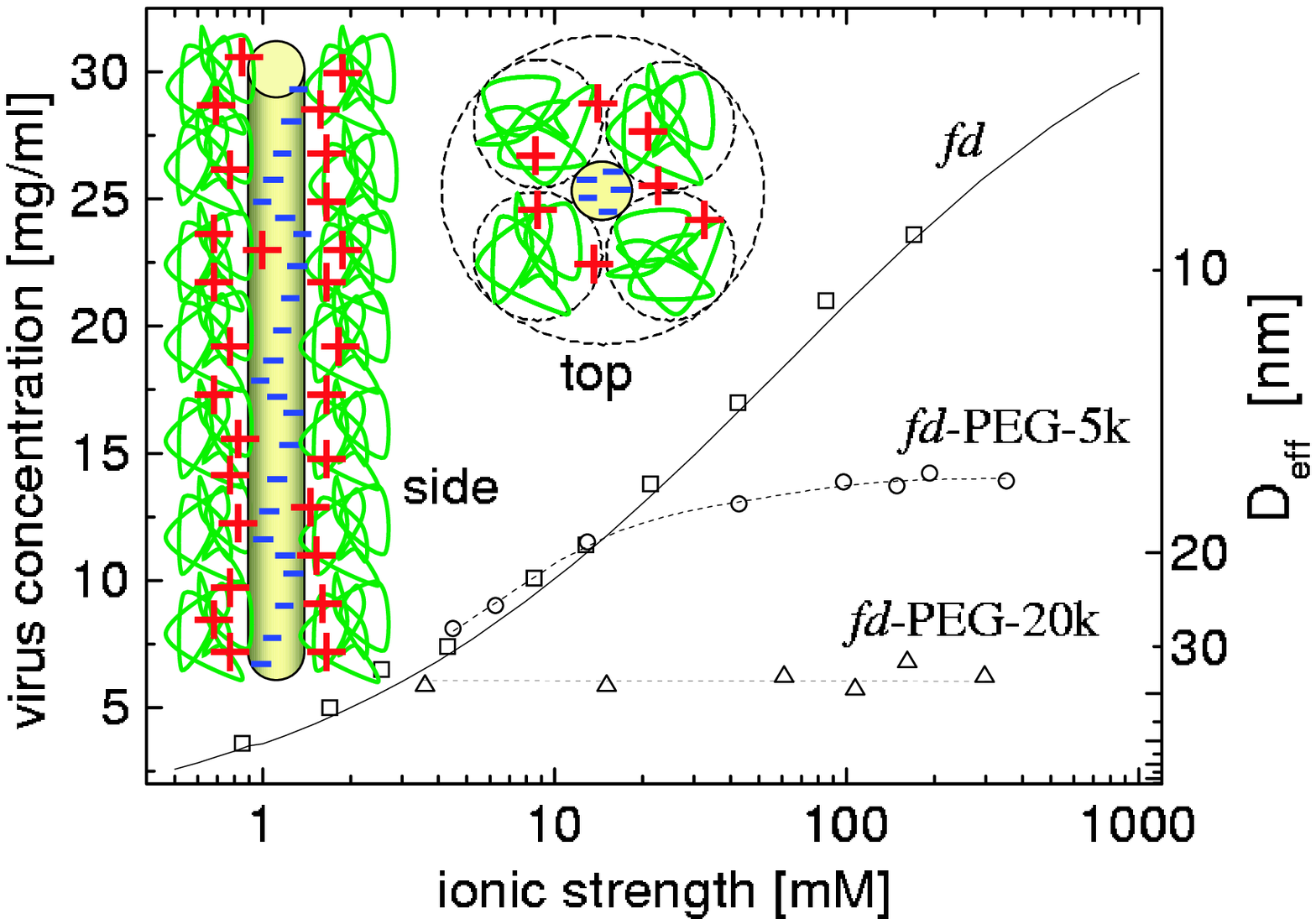,width=85mm}}
\caption{Isotropic - cholesteric phase boundary as a function of
ionic strength for bare virus ({\it fd}) and for virus coated with
neutral polymer
({\it{fd}}-PEG-5k and {\it{fd}}-PEG-20k). 
The lines represent the highest concentration for which the
isotropic phase, I, is stable (below line), whereas the range of
the cholesteric phase, N*, is above each line. Squares and circles
indicate the I-N* transition in {\it fd} and in {\it fd} coated
with PEG-5k respectively, while triangles refer to the {\it fd}
virus coated with PEG-20k. The concentration of the I-N*
transition for bare {\it fd} increases as a function of ionic
strength as 1/$D_{\mbox{\scriptsize eff}}$ (solid line), where
$D_{\mbox{\scriptsize eff}}$ is the effective diameter taking into
account the electrostatic repulsion between rods at the I-N*
transition \cite{Onsager}. {\it fd} with grafted PEG shows
identical behavior to non-grafted, up to an ionic strength
corresponding to a $D_{\mbox{\scriptsize eff}}$ that matches the
size of the stabilizing polymer layer. For a higher amount of
added salt, the phase diagram becomes independent of ionic
strength, and the {\it{fd}}-PEG system behaves as a sterically
stabilized suspension. This is schematically represented by the
cartoon of a virus with polymers grafted on its surface where
$D_{\mbox{\scriptsize eff}}^{\mbox{\scriptsize electrostatic}}<
 D_{\mbox{\scriptsize eff}}^{\mbox{\scriptsize polymer}}$.
} \label{PhaseDiagram}
\end{figure}

Although the {\it fd} virus is charged in solution, it is possible
to prepare 
sterically stabilized colloids by irreversibly binding 
neutral polyethylene glycol (PEG) of different molecular weights
(5,000 or 20,000 g/Mol) to the surface of the virus
\cite{RoyalSoc}. End functionalized PEG molecules 
(Shearwater corp.) have been used to covalently attach the water
soluble polymer to the amino groups on the coat proteins of {\it
fd} that are exposed to the solution. The refractive index
increment, dn/dc, has been measured to estimate the degree of
coverage of {\it fd}-PEG virus \cite{dn/dc}, and 195$\pm$30
polymers PEG-20k (400$\pm$40 for PEG-5k) are attached to each
virus. This result corresponds to nearly complete coverage of the
virus by the polymer. Indeed, by considering grafted polymers as
spheres of radius of gyration R$_{g}$, about $ \pi
(1+D_{\mbox{\scriptsize bare}}/{2R_{g}})\times L / {2R_{g}} $
polymers can be close-packed around a virus of diameter
$D_{\mbox{\scriptsize bare}}$ (Fig. \ref{PhaseDiagram}).
In the case of PEG-20K (R$_{g}$= 7~nm \cite{RgPEG}), this model
predicts about 250 grafted polymers per virus, which is close to
measured value.

The isotropic - cholesteric (I-N*) transition has been
investigated for this system of {\it fd} virus grafted with PEG
\cite{RoyalSoc}.
According to Onsager \cite{Onsager}, an isotropic to nematic phase
transition occurs in suspensions of hard rods when the rod number
density $n$ reaches $n \pi L^2 D \simeq 16$ (eq. 1). This is
experimentally observed for bare {\it fd} (without any grafted
polymer) over a wide range of ionic strengths as shown with square
symbols in Fig. \ref{PhaseDiagram}. Note that there is no
difference in Onsager's theory
between the isotropic-nematic and isotropic-cholesteric phase
transitions. The free energy difference between the nematic and
cholesteric phase is negligible compared to the free energy
difference between the isotropic and nematic phase \cite{Saupe}.

Because the {\it fd} virus is charged, it is necessary to account
for electrostatic repulsion by considering the colloids as having
an effective diameter ($D_{\mbox{\scriptsize eff}}$) larger than
their bare diameter ($D_{\mbox{\scriptsize bare}}$) 
with $D_{\mbox{\scriptsize eff}}$ roughly proportional to the
Debye screening length \cite{Onsager}. $D_{\mbox{\scriptsize
eff}}$ therefore decreases with increasing ionic strength
\cite{Tang95} and
approaches $D_{\mbox{\scriptsize bare}}$ at high ionic strength.
By substituting $D_{\mbox{\scriptsize bare}}$ with
$D_{\mbox{\scriptsize eff}}$ in equation (1), it follows that the
concentration of 
{\it fd} at the I-N* transition is inversely proportional to
$D_{\mbox{\scriptsize eff}}$. For {\it fd}-PEG rods, there is an
ionic strength at which the coexistence concentrations of the I-N*
transition become independent of ionic strength. The crossover
from the regime where the rods are electrostatically stabilized to
where they are sterically stabilized by ``soft'' repulsion between
virus-bound polymers is shown in Fig. \ref{PhaseDiagram}. This
corresponds to a ionic strength of I=~3~mM for {\it fd}-PEG-20k
and I=~20~mM for {\it fd}-PEG-5k. At this ionic strength, the
diameter associated with electrostatic interactions
($D_{\mbox{\scriptsize eff}}$) equals the diameter of the virus
plus grafted polymer. It is therefore possible to extract an
effective diameter for the polymer coated virus. In this range of
ionic strengths, between 15 and 200 mM, the polymer is in a good
solvent and the effects of NaCl on PEG properties are negligible
in terms of radius of gyration and viscosity \cite{bailey}. From
Fig. \ref{PhaseDiagram}, the effective diameter of {\it
fd}-PEG-20k at I= 3~mM is 37~nm, which is approximately equal to
the hard diameter of {\it fd} ($D_{\mbox{\scriptsize bare}}$=
6.6~nm) plus twice the diameter of PEG ($R_g$= 7~nm \cite{RgPEG}):
$D_{\mbox{\scriptsize bare}}+4R_g=35$~nm. The {\it fd}-PEG-5k
complex has $D_{\mbox{\scriptsize eff}} = 17$~nm at high ionic
strength, while $D_{\mbox{\scriptsize bare}}+4R_g = 19$~nm ($R_g$=
3~nm for PEG-5k \cite{RgPEG}).

\begin{figure}[tbh]
\centerline{\epsfig{file=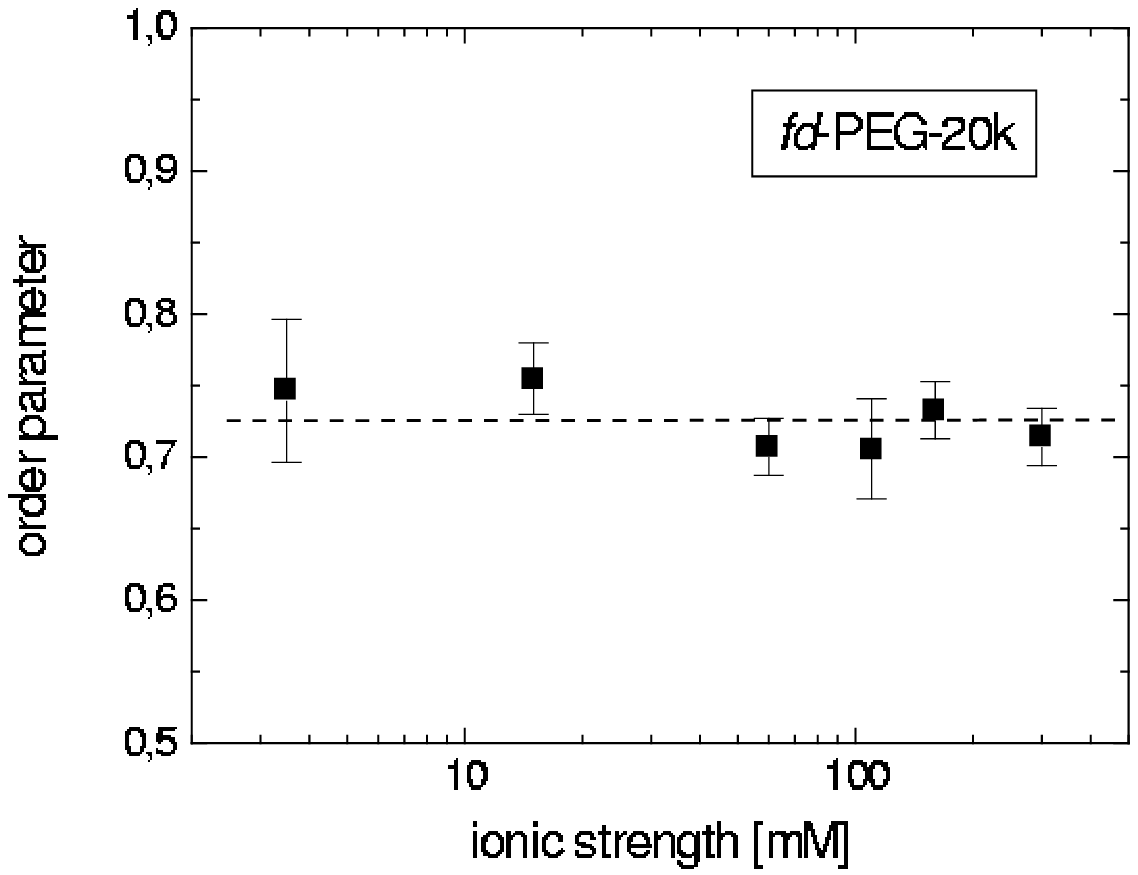,width=70mm}}
\caption{Ionic strength dependence of the nematic order parameter
of {\it fd} virus coated with PEG-20k measured close to the I-N*
transition (samples concentrations around c$\sim$ 9 mg/ml). The
order parameter is obtained from birefringence measurements after
unwinding the cholesteric phase with a magnetic field. The dashed
line is a guide for the eyes.} \label{OrderPara}
\end{figure}
\begin{figure}[tbh]
\centerline{\epsfig{file=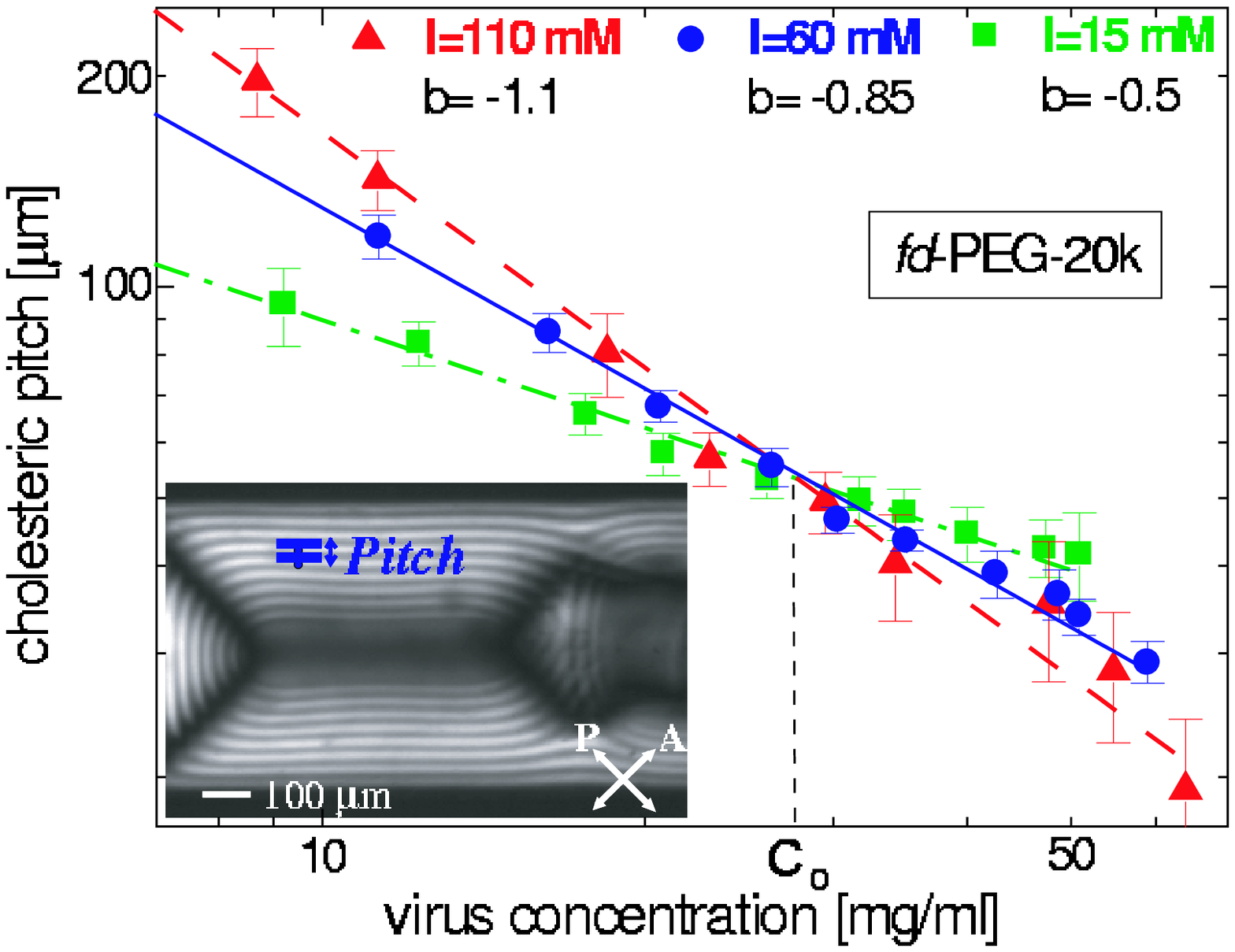,width=85mm}}
\caption{Evolution of the cholesteric pitch (P) with the
concentration (c) of {\it fd} virus grafted with PEG-20k, for
three different ionic strengths (I=15, 60, and 110 mM).
Cholesteric pitch was measured from the ``fingerprint'' texture,
as shown in the insert. The sample is contained in a cylindrical
glass capillary of diameter 0.7 mm and several cm long. Error bars
have been obtained by repeating the same measurement along the
full length of the sample. The scaling exponent of the fit to
$P\propto c^{b}$ is indicated for each ionic strength.}
\label{pitchFd-PEG}
\end{figure}

The good agreement between the measured and estimated value of the
{\it fd}-PEG effective diameter supports the model of polymer
behaving mainly as a sphere 
attached to the surface of the virus. 
Thus, the independence of the phase diagram on ionic strength
reported in Fig. \ref{PhaseDiagram} shows that {\it fd} virus
coated with PEG-20k is a system whose stability is entirely
determined by polymer repulsion. 
This result has been confirmed by measuring the nematic order
parameter of {\it fd}-PEG-20k  close to the I-N* transition. Fig.
\ref{OrderPara} indicates that the order parameter, determined by
birefringence under magnetic field, is independent of ionic
strength just as is the phase diagram 
\cite{RXkirstin}.
Therefore, {\it fd}-PEG-20k liquid crystal turns out to be a
sterically stabilized system, that, surprisingly, continues to
exhibit a cholesteric phase as shown in Fig. \ref{pitchFd-PEG}.
Indeed, one could expect that grafted polymers on virus surface
would screen chiral interactions, leading thus only to a nematic
phase. To our knowledge, {\it fd} virus coated with PEG is the
first completely sterically stabilized system which still has
chiral features. 

The insert of Fig. \ref{pitchFd-PEG} presents a typical
``fingerprint'' texture of {\it fd}-PEG suspensions characteristic
of a cholesteric phase observed by optical microscopy between
crossed polarizers. After a few days of equilibration following
the preparation of the samples, we have measured the dependence of
cholesteric pitch on concentration and ionic strength (Fig.
\ref{pitchFd-PEG}). Paradoxically, {\it fd}-PEG-20k liquid
crystals have a cholesteric pitch which depends on ionic strength.
Close to the I-N* transition where both the phase diagram and
order parameter are independent of ionic strength (Figs.
\ref{PhaseDiagram} and \ref{OrderPara}), the pitch varies by a
factor 2 from 100 $\mu$m at I=15 mM to 200 $\mu$m at I=110 mM. In
this range of low rod concentrations the mean interparticle
distance ($d_{\mbox{\scriptsize inter}}$) is about 60~nm: the {\it
fd}-PEG surfaces are then separated by typically 3
bare diameters \cite{RXkirstin}. The corresponding volume fraction
at the I-N* transition is $\Phi$=0.2 where $\Phi$ is defined by $\Phi
= c \: N{_A} L \: \pi (D_{\mbox{\scriptsize eff}}/2){^2} / \:
M{_W}$ with $N{_A}$ being the Avogadro's number. We can then
calculate the concentration (c*) for which the grafted polymers of
{\it fd}-PEG-20k begin to overlap (corresponding to $\Phi$*=1):
this
gives c*$\simeq \:$35~mg/ml, 
which is close to the concentration c$_0$=28~mg/ml where the pitch is
independent of ionic strength (Fig. \ref{pitchFd-PEG}).
The concentration c* separates two regimes: we
mainly focus on the dilute regime (c$<$c*), where chiral rods do
not overlap. Complexity increases for c$>$c* because of
short-range interactions between the polymers as well as
electrostatic interactions between the viruses, which can induce
intricate behaviors similar to those seen for instance in DNA
condensation
\cite{PhysicsToday}.

What then is the mechanism of transmission of chiral interactions
between sterically stabilized viruses? The first model of the
origin of intermolecular twist was proposed by Straley who considered two
screw-like molecules with excluded-volume
interactions \cite{Straley}. 
He predicted the pitch to be independent of the concentration,
which is not observed experimentally.
Moreover, if chirality is transmitted by some microscopic detail
of the virus, we then expect that the chiral effects will be
diminished for {\it fd} grafted with polymers. However, the same
range of pitch has been seen for the {\it fd}-PEG system as for
bare {\it fd} (between 10 and 200 $\mu$m) \cite{langmuir}. That
seems to exclude the strict steric model of packing of screws,
because the thread of the screws of {\it fd} virus is covered by
a 30 nm thick polymer coat.
Even with the assumption that the polymers grafted to the helical
protein coat replicate the chiral symmetry of the
underlying bare virus, 
it does not explain why the cholesteric pitch should depend on
ionic strength.

\begin{figure}[tbh]
\centerline{\epsfig{file=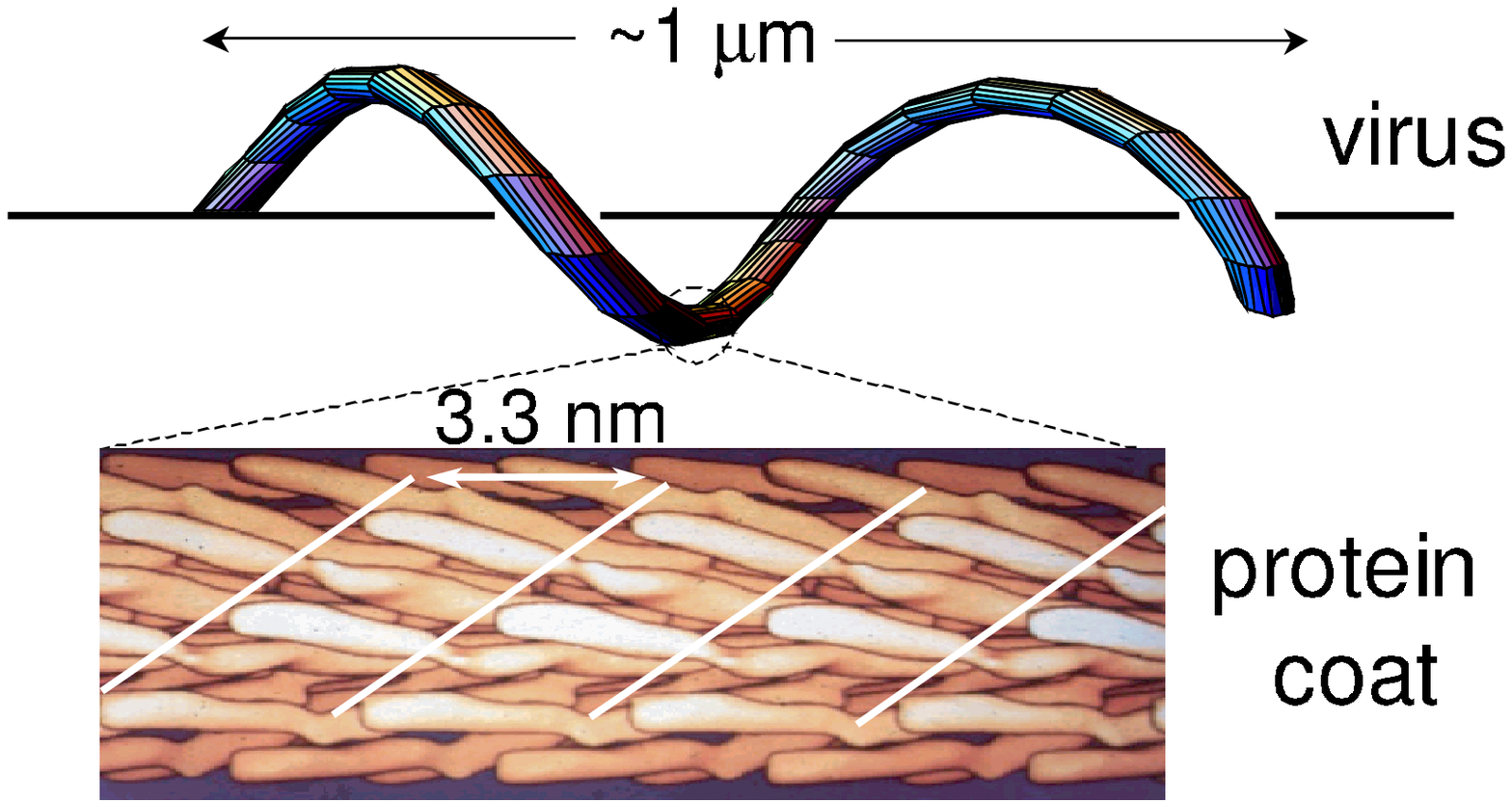,width=85mm}}
\caption{Schematic representation of the ``cork-screw'' model: the
chirality is transmitted between viruses by the helical shape
(accentuated for illustration) of
the whole virus 
at the micron length scale and not by the helical structure of the
protein coat.} 
\label{cork-screw}
\end{figure}

If the chirality is not transmitted at the microscopic length
scale of the helical periodicity of the protein coat (with an
axial repeat of 3.3~nm for {\it fd} virus \cite{makowski}), the
existence of a {\it superhelical} twist, where chirality occurs at
a much larger length
scale, may be proposed.  
Brownian motion causes the virus to bend back and forth, and due
to the coupling between flexibility and the underlying structural
chirality imposed by the helical arrangement of coat proteins, the
virus could twist more in one direction than in another leading to
the creation of a ``cork-screw'' shape with a helical pitch of
order a micron. 
A test [suggested by F. Amblard] of this model would be to measure
the distribution of angles between the tangent vectors of the ends
of a single virus confined to a surface. This is a Gaussian
distribution for the worm-like chain model, but will have a
maximum at a non-zero angle for a bent molecule, such as a
cork-screw of length less than one full pitch. However, the
``cork-screw'' model can not be the whole story as Pf1, a chiral
and flexible virus with a structure very similar to fd, has only a
nematic phase, which implies that not all chiral flexible
molecules form cork-screws. Additionally, completely rigid rods
like cellulose \cite{Revol} and chitin \cite{Revol2} form a
cholesteric phase implying that the cork-screw distortion is not
necessary for cholesteric production. An interesting experiment
would be to coat these rigid molecules with long polymer. If they
still form the cholesteric phase then the cork-screw model would
be rejected. 


\begin{figure}[tbh]
\centerline{\epsfig{file=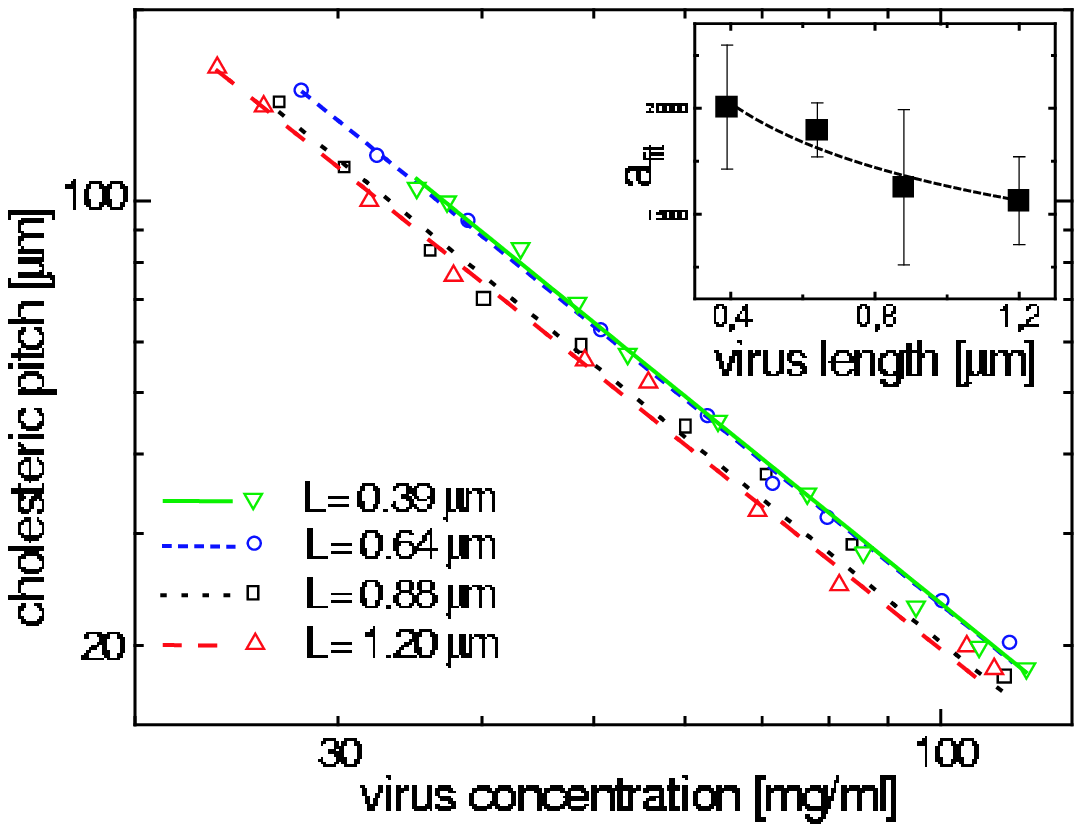,width=85mm}} \caption{Graph
of the cholesteric pitch (P) versus the concentration (c) for
mutant viruses of different contour lengths (L). The ionic
strength is 60 mM. The result of the fit $P=a_{fit}c^{b}$
indicates the same scaling
exponent b=-1.45 for all mutants. 
The insert indicates the dependence of the pitch with the length
of the virus at a given concentration: $P\propto a_{fit} \propto
1/L^{0.25}$. This result contradicts the main existing theories of
chirality in liquid
crystals which predict $P \propto L^2$
\cite{leikin}\cite{Harris}. } \label{mutants}
\end{figure}

One of the most advanced theories of cholesteric assemblies of
charged rods has been proposed by Kornyshev {\it{et al.}} for DNA.
They suggest that various details of the helical symmetry of
macromolecules surface charge pattern may be responsible for
macroscopic twist \cite{leikin}. The main idea is based on direct
chiral electrostatic interaction between long helical molecules.
Their attempt 
is promising, but how can such a theory account, even
qualitatively, for our results on {\it fd} virus stabilized with
polymer? Moreover, the prediction of a cholesteric pitch {\it
increasing} with the square of the viral contour length ($P\propto
L^{2}$) shown both by Kornyshev {\it{et al.}} \cite{leikin} and by
Harris {\it{et al.}} \cite{Harris} is in contradiction to our
results obtained with mutant viruses (without any polymer
grafted). Indeed, we are able to grow monodisperse mutants
\cite{RoyalSoc} where only the contour length varies, keeping
unchanged the local structure of the viruses 
\cite{mutants}. We find
that the cholesteric pitch {\it decreases} with increasing viral
contour length as shown in Fig. \ref{mutants}. These experiments
suggest that current theories do not capture all the essential
elements of the connection between microscopic and macroscopic
chirality in colloidal liquid crystals.


In conclusion, we have presented in this Letter a system of chiral
colloidal rods coated with neutral polymer, which behaves as a
sterically stabilized suspension because the phase diagram and
order parameter are independent of ionic strength. However, this
liquid crystal has the baffling feature that it exhibits a
cholesteric phase whose pitch depends on ionic strength. If the
cholesteric phase is induced by the charges on the the virus which
are buried by a polymer coat of thickness corresponding to several
Debye screening lengths, then we conclude that the cholesteric
phase arises from very weak and subtle forces. Our result
highlights the fundamental role of counterions in the existence of
a macroscopic chirality and should stimulate new theoretical
investigations on this fundamental topic of the origin of
chirality in liquid crystalline phases.

This research was supported by NSF/DMR grants. We acknowledge
Kirstin Purdy for her assistance in the preparation of virus
mutants.


\clearpage

\end{document}